\documentclass[11pt,a4paper]{article}

\usepackage{epsfig,amsmath,amssymb,cite}
\def\Om{\Omega}
\def\om{\omega}
\def\lp{\left(}
\def\rp{\right)}
\def\lb{\left[}
\def\rb{\right]}
\def\be{\begin{equation}}
\def\ee{\end{equation}}
\begin{document}

\title{Brans--Dicke cylindrical wormholes} 
\author{Ernesto F. Eiroa$^{1,2,}$\thanks{e-mail: eiroa@iafe.uba.ar}, 
Claudio Simeone$^{2,3,}$\thanks{e-mail: csimeone@df.uba.ar}\\
{\small $^1$ Instituto de Astronom\'{\i}a y F\'{\i}sica del Espacio, C.C. 67, 
Suc. 28, 1428, Buenos Aires, Argentina}\\
{\small $^2$ Departamento de F\'{\i}sica, Facultad de Ciencias Exactas y 
Naturales,} \\ 
{\small Universidad de Buenos Aires, Ciudad Universitaria Pab. I, 1428, 
Buenos Aires, Argentina} \\
{\small $^3$ IFIBA, CONICET, Ciudad Universitaria Pab. I, 1428, 
Buenos Aires, Argentina}} 
\maketitle

\begin{abstract}

Static axisymmetric thin-shell wormholes are constructed within the framework of the  Brans--Dicke scalar-tensor theory of gravity. Examples of wormholes associated with vacuum and electromagnetic fields are studied. All constructions must be threaded by exotic matter, except in the case of geometries  with a singularity of finite radius, associated with an electric field, which can have a throat supported by ordinary matter. These results are achieved with any of the two definitions of the flare-out condition considered.\\

\noindent 
PACS number(s):  04.20.Gz, 04.50.Kd, 04.40.Nr

\end{abstract}

\section{Introduction}

The study of wormholes began in the early days of general relativity \cite{old} (for a historical perspective see Ref. \cite{visser}). After traversable Lorentzian wormholes were introduced by Morris and Thorne \cite{mo}, such topologically non trivial solutions of the equations of gravitation have received considerable attention \cite{visser}. However, the necessity of exotic matter (matter not fulfilling the energy conditions) supporting compact wormhole geometries, which seems unavoidable within the framework of general relativity, constitutes the main obstacle for the actual existence of these objects. In particular,  this  aspect has been analyzed in detail for wormholes of the thin-shell class \cite{povi,visser}, that is, configurations which are mathematically constructed by cutting and pasting two geometries, so that a matter layer is located at the throat (see Ref. \cite{ernesto} and references therein). 

In the last two decades, cylindrically symmetric geometries have deserved a considerable attention, mainly in relation to cosmic strings. These have been the object of a detailed study, because of the important role they could have played in structure formation in the early universe, and also for their possible observation  by gravitational lensing effects (see Ref. \cite{vilenkin}). Besides,  open or closed fundamental strings are the center of present day attempts towards a unified theory. Then the interest in the gravitational effects of both fundamental and cosmic strings, and in general in axisymmetric solutions, has been recently  renewed   (for example, see Refs. \cite{strings}). As a natural consequence, cylindrically symmetric wormholes  have been studied  in the last few years; see Refs. \cite{cle,brle,cil,eisi10}.  It was noted in Ref. \cite{brle} that for non compact wormholes --as cylindrically symmetric ones are-- there are two admissible definitions of the throat, and that one of them --that the geodesics {\it restricted to a plane normal to the symmetry axis} open up-- can be compatible in general relativity with the energy conditions. These cylindrical solutions satisfying the weak energy condition cannot have a flat or string asymptotic behavior at both sides of the throat. In particular, non exotic wormholes associated with an azimuthal magnetic field were found, which are neither symmetric with respect to the throat nor flat or string like at infinity (see Ref. \cite{brle} and references therein). Thin-shell cylindrical wormholes with positive energy density at the throat were found in Ref. \cite{eisi10} within the framework of Einstein's gravity.

On the other hand, the issue of matter supporting wormholes has been revisited in alternative theories, and it was shown that the requirement of exotic matter could in some cases  be avoided \cite{gra-wi,nts}. In particular, some years ago,  Anchordoqui {\it et al.} \cite{ancho} showed that, in Brans--Dicke gravity,  Lorentzian  wormholes of the Morris--Thorne type are compatible with matter which, apart from the Brans--Dicke scalar field, satisfies the energy conditions; an analogous result was found in \cite{ers} for spherical thin-shell configurations. Other related aspects of wormholes in Brans-Dicke or in scalar-tensor theories were also discussed in Refs. \cite{agnese,other,brst}. 

In the present article we address the construction of static cylindrical thin-shell wormholes in Brans--Dicke theory, and we study in detail the energy conditions for the matter supporting these objects. We work with the two definitions of the flare-out condition proposed for axisymmetric configurations.  We consider some examples associated with vacuum and Brans--Dicke--Maxwell cylindrical spacetimes. Units such as $G=c=1$ are adopted.

\section{Wormhole construction}

In Brans--Dicke theory,  matter and non-gravitational fields generate a long range scalar field $\phi$ which, together with them, acts as a source of gravitational field. The metric equations generalizing those of general relativity are, in the Jordan frame, 
\be
R_{\mu\nu}-\frac{1}{2}g_{\mu\nu}R=\frac{8\pi}{\phi} T_{\mu\nu}+\frac{\omega}{\phi^2}\phi_{,\mu}\phi_{,\nu}-\frac{\omega} {2\phi^2} g_{\mu\nu}\phi_{,\alpha}\phi^{,\alpha}+\frac{1}{\phi}\phi_{;\mu;\nu}-\frac{1}{\phi}g_{\mu\nu} \phi_{:\alpha}^{;\alpha},\label{ebd1}
\ee 
where $R_{\mu\nu}$ is the Ricci tensor, $T_{\mu\nu}$ is the  energy-momentum tensor of  matter and fields (not including the  Brans--Dicke  field).
The field $\phi$ is a solution of the equation 
\be
\phi_{;\mu}^{;\mu}=\frac{1}{\sqrt{-g}}\frac{\partial}{\partial x^\mu}\left(\sqrt{-g}\ g^{\mu\nu}\frac{\partial\phi}{\partial x^\nu}\right)=\frac{8\pi T}{3+2\omega},\label{ebd2}
\ee
where $T$ is the trace of $T_{\mu\nu}$. With $\phi=\mathrm{constant}=1$ the Einstein's equations are recovered. The dimensionless constant $\om$ can, in principle, take any value. Its relation to the character of the scalar field is best understood by recalling the Einstein frame Lagrangian \cite{brst}. There it is apparent that $\om+3/2$ determines the sign of the kinetic term of the field $\phi$: when $\om>-3/2$ the field is normal, if $\om<-3/2$ this term is negative and the field is a ghost, and when $\om=-3/2$ the field loses its dynamics. From now on we shall exclude the particular value $\om=-3/2$ in our analysis. 

We start the mathematical construction of a wormhole  from a solution of the field equations (\ref{ebd1}) and (\ref{ebd2}). The most general static metric with cylindrical symmetry has the form
\begin{equation}
ds^2 = -A(r)dt^2 +B(r)dr^2 +C(r)d\varphi ^2+D(r)dz^2,
\label{e1}
\end{equation}
where $A$, $B$, $C$ and $D$ are positive functions of $r$. From this static geometry we take two copies $ \mathcal{M}^{\pm} = \{ x / r \geq a \}$ of the region defined by  $r \geq a$ and join them at the hypersurface given as
$ \Sigma \equiv \Sigma^{\pm} = \{ x / r - a = 0 \}$. Thus we have  a new, geodesically complete, manifold $\mathcal{M}=\mathcal{M}^{+} \cup \mathcal{M}^{-}$.
The throat of the wormhole is a synchronous timelike hypersurface, in which we adopt coordinates $\xi ^i=(\tau , \varphi,z )$, with $\tau $ the proper time on the matter shell placed at $r=a$. The components of the extrinsic curvature  corresponding to the two sides of the shell are
\begin{equation}
  K_{ij}^{\pm} = - n_{\gamma}^{\pm} \left. \left( \frac{\partial^2
  X^{\gamma}}{\partial \xi^i \partial \xi^j} +\Gamma_{\alpha\beta}^{\gamma}
  \frac{\partial X^{\alpha}}{\partial \xi^i} \frac{\partial
  X^{\beta}}{\partial \xi^j} \right) \right|_{\Sigma}, \label{e2}
\end{equation}
where $n_{\gamma}^{\pm}$ are the unit normals ($n^{\gamma} n_{\gamma} = 1$) to
$\Sigma$ in $\mathcal{M}$, which with the definition ${\cal H}(r)=r-a=0$ read:
\begin{equation}
  n_{\gamma}^{\pm} = \pm \left| g^{\alpha \beta} \frac{\partial
  \mathcal{H}}{\partial X^{\alpha}} \frac{\partial \mathcal{H}}{\partial
  X^{\beta}} \right|^{- 1 / 2} \frac{\partial \mathcal{H}}{\partial
  X^{\gamma}} . \label{e3}
\end{equation}
If the convenient  orthonormal basis $\{ e_{\hat{\tau}} = \sqrt{1/A(r)}e_{t}$, $e_{\hat{\varphi}} 
= \sqrt{1/C(r)}e_{\varphi}$, $e_{\hat{z}} =\sqrt{1/D(r)}e_{z}\}$ 
is adopted on the shell, then the induced $2+1$ metric reads $g_{_{\hat{\imath} \hat{\jmath}}} = \eta_{_{\hat{\imath}
\hat{\jmath}}}=diag(-1,1,1)$, and  with this choice, for the metric (\ref{e1})  we obtain
\begin{equation}
K_{\hat{\tau} \hat{\tau}}^{\pm} = \mp \frac{A'(a)
}{2A(a) \sqrt{B(a)}},
\label{e4}
\end{equation}
\begin{equation}
K_{\hat{\varphi} \hat{\varphi}}^{\pm} = \pm \frac{C'(a)}{2C(a)\sqrt{B(a)}}, 
\label{e5}
\end{equation}
and
\begin{equation}
K_{\hat{z} \hat{z}}^{\pm} = \pm \frac{D'(a)}
{2D(a) \sqrt{B(a)}}.
\label{e6}
\end{equation}
We have used the usual notation in which  the prime represents  $d/dr$. The Brans--Dicke  equations on the shell, that is the Lanczos equations \cite{daris,mus} for Brans--Dicke gravity, take the form \cite{dahia}
\begin{equation}
-[K_{\hat{\imath} \hat{\jmath}}]+[K]g_{\hat{\imath} \hat{\jmath}}=
\frac {8\pi}{\phi} \left(S_{\hat{\imath} \hat{\jmath}}-\frac{S}{2\omega +3}\,g_{\hat{\imath} \hat{\jmath}}\right),
\label{e7}
\end{equation}
\be
[\phi_{,N}]=\frac{8\pi S}{2\omega+3},\label{deltaphi}
\ee
where, as usual, $[K_{_{\hat{\imath} \hat{\jmath}}}]\equiv K_{_{\hat{\imath}
\hat{\jmath}}}^+ - K_{_{\hat{\imath} \hat{\jmath}}}^-$, 
$[K]=g^{\hat{\imath} \hat{\jmath}}[K_{\hat{\imath} \hat{\jmath}}]$ is the 
trace of $[K_{\hat{\imath} \hat{\jmath}}]$ and
$S_{_{\hat{\imath} \hat{\jmath}}} = \text{\textrm{diag}} ( \sigma, 
p_{\varphi }, p_{z} )$ is the surface stress-energy tensor of all matter and fields apart from the Brans--Dicke field. This field is continuous across the surface $\Sigma$, but its normal derivative $\phi_{,N}=\phi_{,\alpha}n^{\alpha}$ has a jump $[\phi_{,N}]$ associated with the trace $S$ of the induced surface stress-energy tensor.  By replacing Eqs. (\ref{e4}), (\ref{e5}) and (\ref{e6}) in Eq. (\ref{e7}) and taking into account Eq. (\ref{deltaphi}) we have that the energy density and pressures for a static configuration are given by  
\begin{equation}
\sigma = - \frac{\phi(a)}{8 \pi \sqrt{B(a)}}  \left[
\frac{C'(a)}{C(a)} + \frac{D'(a)}{D(a)} +2\frac{\phi'(a)}{\phi(a)}\right],
\label{e12}
\end{equation}
\begin{equation}
p_{\varphi} =  \frac{\phi(a)}{8 \pi \sqrt{B(a)} } 
 \left[ \frac{A'(a)}{A(a)} + \frac{D'(a)}{D(a)} + 2\frac{\phi'(a)}{\phi(a)}\right],
\label{e13}
\end{equation}
\begin{equation}
p_{z} =  \frac{\phi(a)}{8 \pi \sqrt{B(a)}} 
\left[\frac{A'(a)}{A(a)} + \frac{C'(a)}{C(a)} + 2\frac{\phi'(a)}{\phi(a)}\right] .
\label{e14}
\end{equation}
In order to satisfy Eq. (\ref{deltaphi}), the field $\phi$ and the metric functions must fulfill
\be
2\om\frac{\phi'(a)}{\phi(a)}=\frac{A'(a)}{A(a)}+\frac{C'(a)}{C(a)}+\frac{D'(a)}{D(a)},\label{phiprima}
\ee
that imposes a constraint on the possible values of the throat radius $a$ for which the wormhole construction can be made from a given metric.  The definition of the wormhole throat  usually adopted for compact configurations characterizes it as a {\it minimal area surface}. In the case of cylindrical geometries, defining the  area function ${\cal A}(r)=\sqrt{C(r)D(r)}$, this condition implies that ${\cal A}$ should increase at both sides of the throat, which means  ${\cal A}'(a)>0$, so that $\lp CD \rp'(a)>0$. This can be called  the {\it areal} flare-out condition. In general relativity the scalar field is not present in the junction conditions, and if $\lp CD \rp'(a)=C'(a)D(a)+C(a)D'(a)>0$, from Eq. (\ref{e12}) with $\phi=1$ and $\phi'=0$ one obtains that  the surface energy density is negative, so that matter at the throat is \textit{exotic}. But this is not necessarily the case  in Brans--Dicke gravity because of the term associated with the field $\phi$. 

A different definition of the flare-out condition has been recently  proposed  for infinite cylindrical configurations\cite{brle}. Because a wormhole is defined by its non trivial topology, which constitutes a global property of spacetime, and   for  static cylindrically symmetric geometries the global properties are determined by the behavior of the circular radius function  ${\cal R}(r)=\sqrt{C(r)}$, then the flare-out condition has been defined in Ref.\cite{brle} by requiring that this function  ${\cal R}(r)$ has a minimum at the throat. This definition can be called the {\it radial} flare-out condition, and implies $C'(a)>0$. In our recent work \cite{eisi10} we found that with this definition and in general relativity, a positive energy density is compatible with the existence of cylindrical thin-shell wormholes (though the energy conditions are not satisfied in the examples studied).

With respect to the global properties of wormholes, besides the existence of a throat, a certain asymptotical behavior could also be required. In the case of cylindrical wormholes, a desirable asymptotics would be flat or cosmic string like \cite{brle}. A less restrictive condition could be the existence of spatial infinities \cite{brst}. Though, wormhole topologies have also been considered for which the throat connects two regions that end at a finite distance.

\section{Energy conditions}

Within the framework of Einstein's theory of gravity, a general property of compact wormholes is the requirement of exotic matter, that is, matter which does not fulfill the energy conditions. The weak energy condition (WEC) states that for any timelike vector $u^\mu$ the energy-momentum tensor must satisfy $T_{\mu \nu}u^{\mu }u^{\nu }\ge 0$; the WEC  implies, by continuity, the null energy condition (NEC), i.e. that for any null vector $k^\mu$ it must be $T_{\mu \nu}k^{\mu }k^{\nu }\ge 0$ \cite{visser}. In an orthonormal basis the WEC reads $
\rho \ge 0,\ 
\rho +p_{j}\ge 0\  \forall j$, while the NEC takes the form $
\rho +p_{j}\ge 0\ \forall j$. In our construction we will cut and paste metrics for which the energy conditions are satisfied, so the exotic matter can only be present at the shell.  At the throat we have $p_r=0$, tangential pressures $p_\varphi$ and $p_z$, and surface energy density $\sigma$; thus the conditions to be satisfied by non exotic matter would be  $\sigma \geq0$, $\sigma +p_z\geq 0$ and $\sigma+p_\varphi \geq 0$. For  any of the two ways in which the wormhole throat is characterized  (areal or radial flare-out definitions), in terms of the components of the metric from which we start the construction, the conditions to be examined are 
\be
\sigma=\sigma+p_r= - \frac{\phi(a)}{8 \pi \sqrt{B(a)}} \left[
\frac{C'(a)}{C(a)} + \frac{D'(a)}{D(a)} +2\frac{\phi'(a)}{\phi(a)}\right]\geq 0.\label{smasp}\ee
\be
\sigma+p_\varphi=\frac{\phi(a)}{8\pi\sqrt{B(a)}}\lb\frac{A'(a)}{A(a)}-\frac{C'(a)}{C(a)}\rb\geq 0,\label{pt}
\ee
\be
\sigma+p_z=\frac{\phi(a)}{8\pi\sqrt{B(a)}}\lb\frac{A'(a)}{A(a)}-\frac{D'(a)}{D(a)}\rb\geq 0.\label{pz}
\ee
These inequalities must be analyzed under the restriction imposed by Eq. (\ref{phiprima}). In general, for the class of metrics considered in the examples below, the radial dependence of  the Brans--Dicke field has the power form  $\phi(r)=\phi_0\,r^{1-n}$ so that $\phi'(r)=(1-n)\phi_0\, r^{-n}.$ Therefore the condition (\ref{smasp}) can be recast as
\be
\sigma=\sigma+p_r= - \frac{\phi_0\,a^{1-n}}{8 \pi \sqrt{B(a)}} \left[
\frac{C'(a)}{C(a)} + \frac{D'(a)}{D(a)} +\frac{2(1-n)}{a}\right]\geq 0.\label{smasp2}\ee
If a cylindrical wormhole construction satisfies Eqs. (\ref{smasp}), (\ref{pt}) and (\ref{pz})  
then it is supported by a shell of ordinary matter. We shall address this point in the following examples, in which the wormhole construction starts from vacuum and Brans--Dicke--Maxwell solutions. As we shall see, the most interesting case will turn out to be the one associated with a radial electric field, which allows for a configuration supported by a shell of matter satisfying the energy conditions.

\section{Examples}

\subsection{Vacuum metric}
The general static solution with cylindrical symmetry of the Brans--Dicke equations (\ref{ebd1}) and (\ref{ebd2}) in the vacuum case $T_{\mu\nu}=0$ is given by the metric functions \cite{as00,bd09}
\be
A(r)=B(r)=r^{2d(d-n)+\Om(\om)}, \label{va1}
\ee
\be
 C(r)=W_0^2r^{2(n-d)},\label{va2}
\ee
\be
 D(r)=r^{2d},\label{va3}
\ee
where $n$ and $d$ are constants of integration; $n$ is related to the departure from pure general relativity --see below-- and $d$ can be understood as a mass parameter.  We have introduced the definition\footnote{In Ref. \cite{as00} there is a typo in Eq. (31), where a factor $(n-1)$ multiplying $\om$ is missing, as can be deduced by comparison with the previous equations.} $\Om(\om)=[\om(n-1)+2n](n-1)$. The scalar field takes the form 
\be 
\phi=\phi_0r^{1-n}.\label{campo}
\ee
The solution for Einstein's gravity is obtained for $n=1$ (which corresponds to a uniform field $\phi=\phi_0$), and is the well known Levi--Civita metric. By replacing the metric functions in Eqs. (\ref{e12}), (\ref{e13}) and (\ref{e14}) the energy density and the pressures at the throat are obtained. From Eq. (\ref{phiprima}) the throat radius $a$ is given in terms of the parameters. For the coefficients given by Eqs. (\ref{va1}), (\ref{va2}) and (\ref{va3}) we have $C'(a)/C(a)=2(n-d)/a$ and  $C(a)D(a)=W_0^2a^{2n}$, so that  $(C'/C)(a)+(D'/D)(a)= 2n/a$. Thus the radial flare-out condition requires $n-d>0$, while the areal flare-out condition requires $n>0$. If any of the flare-out conditions is satisfied, the manifold corresponds to a wormhole which extends to infinity at both sides of the throat. Outside the throat there is vacuum, so the energy conditions are fulfilled. By substituting $\phi $ and $\phi'$ in the expressions for the energy density and pressures of the associated thin-shell wormhole, we immediately obtain
\be
\sigma=\sigma+p_r= - \frac{\phi_0}{4\pi a^n\sqrt{B(a)}}.
\ee
The wormhole construction then cannot satisfy the energy conditions at the throat for a positive Brans--Dicke field.

\subsection{Magnetic fields}

The geometries induced by static axisymmetric magnetic fields  have been recently studied in Brans--Dicke theory \cite{bd09}.
The cylindrically symmetric geometry corresponding to an axial static  magnetic field (and thus associated with an angular stationary  current distribution) is determined by the metric functions
\be
A(r)=B(r)=r^{2d(d-n)+\Om(\om)}\lp 1+c^2r^{-2d+n+1}\rp^2,
\ee 
\be
C(r)=W_0^2 r^{2(n-d)}\lp1+c^2r^{-2d+n+1}\rp^{-2},
\ee
\be
 D(r)=r^{2d}\lp1+c^2r^{-2d+n+1}\rp^{2}.
\ee
Instead, for an axial stationary current we have an azimuthal static magnetic field, which has associated the axisymmetric geometry described by 
\be
A(r)=B(r)=r^{2d(d-n)+\Om(\om)}\lp 1+c^2r^{2d-n+1}\rp^2 ,
\ee 
\be
C(r)=W_0^2 r^{2(n-d)}\lp1+c^2r^{2d-n+1}\rp^{2},
\ee
\be
D(r)=r^{2d}\lp1+c^2r^{2d-n+1}\rp^{-2}.
\ee
In both cases $\Omega(\omega)$ is defined as above,  and $d$ and $n$ are constants of integration with a physical meaning analogous to the vacuum case; the constant $c\neq 0$ is related to the magnetic field strength \cite{bd09}. The Brans--Dicke field has the same form as in the vacuum case, shown in   Eq. (\ref{campo}). The Einstein's gravity solutions \cite{ks} are obtained for $n=1$, while the Brans--Dicke vacuum solution is recovered with $c=0$. The energy density and the pressures at the throat are obtained from  Eqs. (\ref{e12}), (\ref{e13}) and (\ref{e14}), and Eq. (\ref{phiprima}) gives the throat radius $a$ in terms of the parameters. If any of the flare-out conditions is fulfilled, we obtain a wormhole which extends to infinity at both sides of the throat. There is a magnetic field outside the throat, so the energy conditions are satisfied there. Though the behavior of the two Brans--Dicke metrics with magnetic field is different, in both cases the product $CD$ is the same and $(C'/C)(a)+(D'/D)(a)= 2n/a$, so we obtain
\be
\sigma=\sigma+p_r=-\frac{\phi_0}{4\pi a^n\sqrt{B(a)}}.
\ee
Hence  a thin-shell wormhole constructed from the geometries associated with both kinds of magnetic field cannot fulfill the energy conditions at the throat for a positive Brans--Dicke field.

\subsection{Radial electric field}

The static metric associated with a radial electrostatic field was also recently obtained in \cite{bd09}. A cylindrically symmetric static charge distribution leads to a radial electric field which induces an axisymmetric geometry described by the metric functions
\be
A(r)= r^{2d}\lp 1-c^2r^{2d-n+1}\rp^{-2},
\ee
\be
C(r)=W_0^2 r^{2(n-d)}\lp 1-c^2r^{2d-n+1}\rp^{2},
\ee
\be
B(r)=D(r)=r^{2d(d-n)+\Om(\om)}\lp1-c^2r^{2d-n+1}\rp^{2},
\ee
where $d$ and $n$ are integration constants with a meaning analogous to the preceding cases, and $c\neq 0$ is associated with the value of the electric field; $\Omega(\om)$ is defined as before. 
The corresponding Brans--Dicke field is given again by Eq. (\ref{campo}). Thus Einstein's solution \cite{ks} corresponds to $n=1$ which gives a uniform scalar field $\phi=\phi_0$. For  $n=1$ and $c=0$ the well known  Levi--Civita  solution is obtained, which can be put in the form pointed out above by means of a suitable coordinate change  \cite{cqg04}. 
When $2d-n+1\neq 0$, the metric is singular at $r_s=c^{-2/(2d-n+1)}$; then this radius should be excluded as a possible throat radius.   The case $2d-n+1=0$ and  $|c|\neq 1$ is equivalent to a global rescaling of the coordinates of the vacuum case $c=0$, so that it presents no interest, while the case  $2d-n+1=0$ and $|c|=1$ has no physical meaning. From now on we shall consider only the case $2d-n+1\neq 0$.
\begin{figure}[t!]
\centering
\includegraphics[width=14cm]{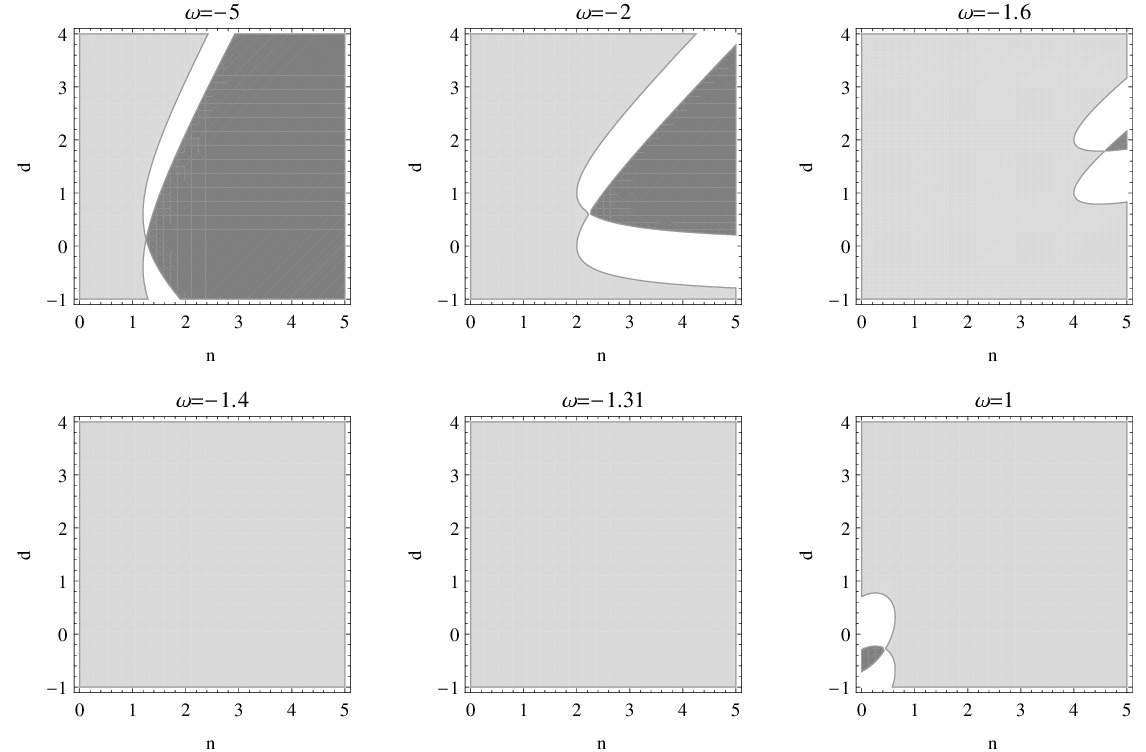}
\caption{Geometries with a radial electric field: The regions where $a>r_s$ (dark gray) and $a<r_s$  (light gray) are shown in terms of the parameters $d$ and $n$, for different values of the Brans--Dicke constant $\om$.  White zones are non physical, since they correspond to values such that the throat radius is not real and positive. The upper row corresponds to values of the Brans--Dicke constant for which the field is a ghost, while the lower row corresponds to normal fields.}
\end{figure}

From the relation (\ref{phiprima}) between the trace of the surface energy-momentum tensor and the jump in the normal derivative of the scalar field we obtain the wormhole throat radius in terms of the parameters and the Brans--Dicke constant:
\begin{equation}
a=\left(\frac{2 d^2 -2 d n +2 n^2 -\omega +n^2 \omega }{c^2 \left(2+4 d+2 d^2-2 n-2 d n+2 n^2- \omega +n^2 \omega \right)}\right)^{1/(1+2 d-n)}.
\label{vinc-elect}
\end{equation} 
The regions $a>r_s$ and $a<r_s$ are shown in Fig. 1. When $a>r_s$  the singular surface is removed from the manifold and the construction is regular and extends to infinity at both sides of the throat, while for $a<r_s$ the geometry is singular at a finite radius. In this case, the configuration is not assymptotically well behaved due to the presence of the singularity; thus the construction could be understood as a wormhole only in the more restricted sense associated with the mere existence of a throat. Taking into account Eq. (\ref{vinc-elect}), the radial flare-out condition can be expressed in the form
\begin{equation}
-2 d - 2 d^2 + 2 d n + 2 n - 2 n^2 + (1-n^2) \omega > 0,\label{FOR}
\end{equation}
and the areal flare-out condition reads
\begin{equation}
-2 d - 2 d^2 + 2 d n - 2 n^2 + (1-n) (3+n) \omega > 0.\label{FOA}
\end{equation}
Outside the shell we have the energy-momentum tensor corresponding to an electric field, so the energy conditions are satisfied there. On the shell, the conditions  $\sigma +p_r \ge 0$, $\sigma +p_{\varphi} \ge 0$, and $\sigma+p_z\geq 0$ lead to
\begin{equation}
-2 + 2 d + 2 d^2 + 2 n - 2 d n + 2 n^2 - (1-n) (3+ n) \omega \geq 0,\label{S+Pr}
\end{equation}
\begin{equation}
 2 d + 2 d^2 - n - 2 d n + 2 n^2 - (1-n^2) \omega \geq 0,\label{S+Pf}
\end{equation}
and
\begin{equation}
2 d + 2 d^2 + 2 n - 2 d n + 2 n^2 - (1-n) (3+ n) \omega  \geq 0,\label{S+Pz}
\end{equation}
respectively. From the equations above we can straightforwardly relate the parameter $n$ with the Brans--Dicke constant: when choosing the radial flare-out condition it is  necessary (but not sufficient) that $\om <-1$ and $n>1$, or $\om >-1$ and $0<n<1$, in order to satisfy  the energy conditions. On the other hand, if the areal flare-out condition is adopted the energy conditions require that $\om <0$ and $n>1$. 
\begin{figure}[t!]
\centering
\includegraphics[width=14cm]{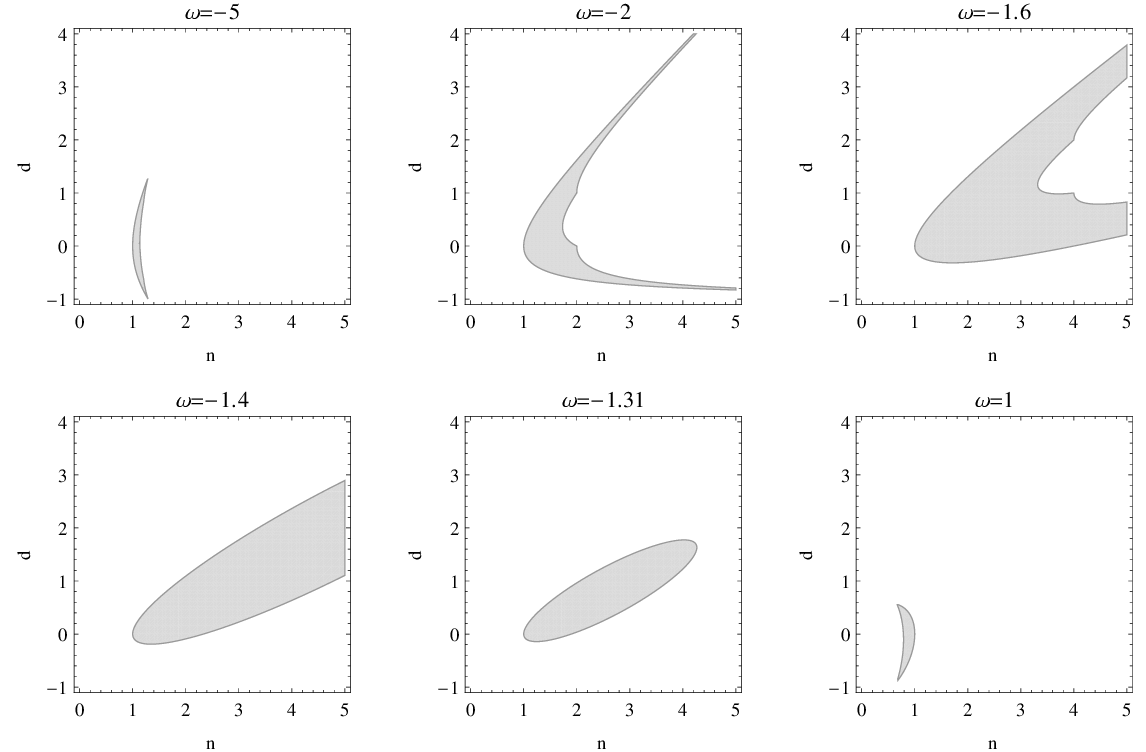}
\caption{Geometries with a radial electric field: The gray zones correspond to values of the parameters such that the energy conditions are satisfied when the radial flare-out condition is adopted.  The upper row corresponds to values of the Brans--Dicke constant for which the field is a ghost, while the lower row corresponds to normal fields.}
\end{figure}
Given the number of parameters involved in the analysis and the range of values that they can take,  a natural approach is to draw the intersection of conditions (\ref{FOR}), (\ref{S+Pr}), (\ref{S+Pf}), and (\ref{S+Pz}) and of the conditions (\ref{FOA}), (\ref{S+Pr}) (\ref{S+Pf}), and (\ref{S+Pz}). The results are shown for the radial flare-out condition in Fig. 2, and for the areal flare-out condition in Fig. 3, choosing different values of the parameters and of the Brans--Dicke constant. By comparing  these figures with Fig. 1, they reveal that in the case of the regular constructions ($a>r_s$), the existence of a throat requires the presence of exotic matter. Instead, for manifolds with a singularity at a finite radius ($a<r_s$), the existence of a throat is compatible with a layer of matter satisfying the energy conditions. This happens with any of both definitions of the flare-out condition and for both normal and ghost scalar fields.
\begin{figure}[t!]
\centering
\includegraphics[width=14cm]{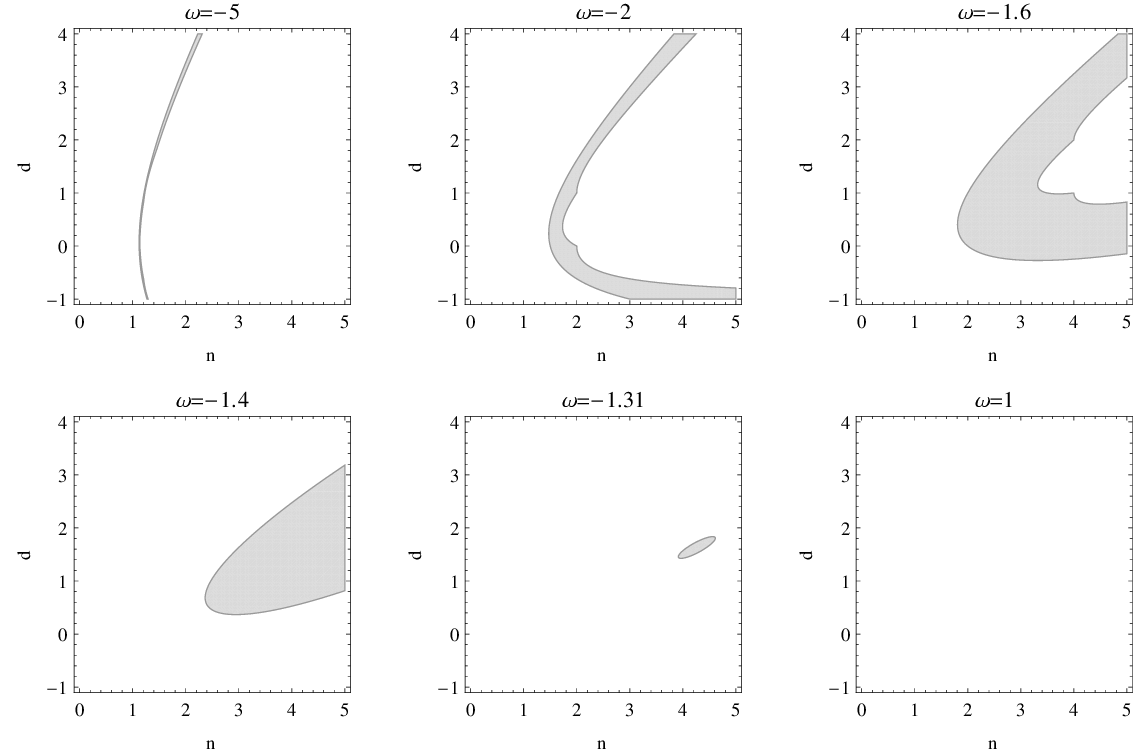}
\caption{Geometries with a radial electric field: The gray zones correspond to values of the parameters such that the energy conditions are satisfied when the areal flare-out condition is adopted.  The upper row corresponds to values of the Brans--Dicke constant for which the field is a ghost, while the lower row corresponds to normal fields.}
\end{figure}

\section{Summary and discussion}\label{discu}

We have constructed static cylindrical thin-shell wormholes within the Brans--Dicke theory of gravity.  Two possible definitions of the flare-out condition --areal and radial-- were considered,  and the character of matter in the shell placed at the throat has been studied.  Examples of wormholes obtained by the cut and paste procedure from the general cylindrical static vacuum solution and from the metrics corresponding to magnetic and electric axisymmetric fields have been examined in detail. In the vacuum and in the magnetic cases the wormhole construction can be performed, but with exotic matter in the shell for all values of the parameters.  In the case associated with a radial electric field, the metric adopted in the construction has a singularity at a finite radius. Wormholes with regular asymptotics can be constructed, but they must be threaded by exotic matter. Configurations with a throat without exotic matter are possible for certain values of the parameters and of the Brans--Dicke constant, but they present a singular surface at a finite radius. This result is obtained for both definitions of the flare-out condition and both normal or ghost Brans--Dicke fields.  The global properties in this case are rather unusual: the geometry opens up at the throat but ends at a singular surface of finite radius (some authors would not call them true wormholes \cite{brst}). The non exotic layers supporting these geometries turn out to be possible even for a non ghost scalar field. In  Refs. \cite{brst} it was shown that wormholes supported by non exotic matter require the presence of ghost fields in scalar-tensor theories of gravity; however, these works deal with wormholes which have compact throats and extend to infinity. We understand that our results regarding throats supported by ordinary matter are a consequence of the non compact character of the configurations considered, and of the ill behavior of the geometry, which has a singularity at a finite radius.

\section*{Acknowledgments}

This work has been supported by Universidad de Buenos Aires and CONICET.

\end{document}